Perspective

Tian-Xiang Zhu, Xiao Liu, Zong-Quan Zhou*, and Chuan-Feng Li

# Remote Quantum Networks based on Quantum Memories

**Abstract:** Quantum networks, capable of transmitting arbitrary quantum states, provide a foundation for a wide range of quantum applications, including distributed quantum computing, distributed quantum sensing, and quantum communication. Photons are the natural carrier of information in quantum networks, but the exponential loss of optical fiber channels prevents the construction of large-scale quantum networks. A potential solution is implementing quantum repeaters based on quantum memories, which can efficiently establish long-distance entanglement from short-distance entanglement. In the past decades, intense efforts have been devoted to constructing large-scale quantum networks based on various atomic quantum memories. In this Perspective, we present a concise overview of current advancements in remote quantum networks, elucidate the imminent challenges that must be addressed, and discuss the future directions.

**Keywords:** Quantum network; Quantum memory; Quantum repeater; Light-matter entanglement

Tian-Xiang Zhu and Xiao Liu contributed equally to this work.

**Tian-Xiang Zhu, Xiao Liu,** CAS Key Laboratory of Quantum Information, University of Science and Technology of China, Hefei 230026, China; Anhui Province Key Laboratory of Quantum Network, University of Science and Technology of China, Hefei 230026, China; CAS Center for Excellence in Quantum Information and Quantum Physics, University of Science and Technology of China, Hefei 230026, China, E-mail: ztx111@ustc.edu.cn (T.-X. Zhu), Xiao.liu@ustc.edu.cn (X. Liu). https://orcid.org/0009-0000-1218-989X (T.-X. Zhu), https://orcid.org/0000-0002-3264-7939 (X. Liu)

**\*Corresponding author: Zong-Quan Zhou,** CAS Key Laboratory of Quantum Information, University of Science and Technology of China, Hefei 230026, China; Anhui Province Key Laboratory of Quantum Network, University of Science and Technology of China, Hefei 230026, China; CAS Center for Excellence in Quantum Information and Quantum Physics, University of Science and Technology of China, Hefei 230026, China; and Hefei National Laboratory, University of Science and Technology of China, Hefei 230088, China, E-mail: zq_zhou@ustc.edu.cn; https://orcid.org/0000-0001-6357-0084

**Chuan-Feng Li,** CAS Key Laboratory of Quantum Information, University of Science and Technology of China, Hefei 230026, China; Anhui Province Key Laboratory of Quantum Network, University of Science and Technology of China, Hefei 230026, China; CAS Center for Excellence in Quantum Information and Quantum Physics, University of Science and Technology of China, Hefei 230026, China; and Hefei National Laboratory, University of Science and Technology of China, Hefei 230088, China, E-mail: cfli@ustc.edu.cn; https://orcid.org/0000-0001-6815-8929

# 1 Introduction

The quantum network represents a significant and intricate objective within the domain of quantum information technology, having attracted considerable attention and investigation over the past decades [1–6]. Quantum networks comprise two principal elements: quantum nodes and quantum channels. Quantum nodes are utilized in processing quantum states, including those employed in quantum computing, quantum communication, quantum sensing, and other applications. Quantum channels are employed for transmitting the quantum states and are characterized by low losses and high transmission fidelity.

Similar to classical networks, photons are the most commonly used information carrier in quantum networks and optical fiber is a widely-used communication channel [7–12]. However, the exponential loss of fiber channels ($\sim$0.2 dB/km for telecom C-band) prevents the direct transmission of single photons over long distances [13]. Unlike classical repeaters that directly amplify signals in classical networks, quantum repeaters face a fundamental challenge: the no-cloning theorem prohibits the perfect cloning of unknown quantum states [14]. The free-space channel [15] exhibits a polynomial loss scaling with distance for extra-atmospheric transmission [16] and can offer lower channel loss over continental distances with satellite links [17, 18], but global coverage and continuous operation still represent great challenges.

To establish quantum networks with arbitrary long distances, H.-J. Briegel et al. proposed a scheme of quantum repeater [19], which employs quantum memories (QMs) and entanglement swapping to reduce the loss scaling in long-distance channels. The core principle is to avoid directly transmitting quantum states over long-distance channels. Instead, the transmission is divided into multiple short-distance elementary links. QMs are used to synchronize the entangled states established at neighboring links. Once they have successfully established entangled states, entanglement swapping would be performed to construct entangle-ment between non-neighboring nodes with longer distances [6, 19–21].

By temporarily storing quantum states, QMs enable the synchronization of probabilistic quantum processes which is a critical capability that not only facilitates the the implementation of quantum repeaters but also supports various important applications in quantum information science. For instance, QMs can improve the secure key rate in quantum key distribution (QKD) protocols [22], deterministically generate multiphoton states [23, 24] for linear optical quantum computing [25, 26] and enable nonlocal quantum gates [27–30] for distributed quantum computing [31–36]. In quantum sensing, the extended storage time of QMs can greatly enhance the sensitivity and overall performance of quantum sensors [37].

QMs can be classified into two principal categories: ensemble-based QMs and single-particle QMs. Ensemble-based QMs leverage the collective interaction between light and an entire atomic ensemble to achieve quantum storage, without the need to track the individual contributions of atoms. This approach benefits from the collective interference effects among many atoms to achieve an efficient light-atom interface. Ensemble-based QMs can easily support multimode storage [38, 39], which is crucial for enhancing the data rate in quantum communication networks. Several physical systems are employed to implement ensemble-based quantum storage, including cold atomic ensembles [8, 11, 40–43], warm atomic vapors [44–48], and rare-earth-ion-doped crystals (REICs) [49–52].

Single-particle QMs, on the other hand, rely on the interaction between photons and individual quantum systems such as single atoms and ions. These systems often require resonant cavities with high-quality factors to enhance the coupling between photons and single-particle systems, ensuring efficient light-matter interaction. A distinct advantage of single-particle QMs is their ability to perform precise control on single atomic qubits in a deterministic way. Moreover, they can serve as ideal sources for single-photon generation, an essential resource for quantum computing and communication. The physical systems used for single-particle quantum storage include single atoms [9, 53–55], trapped ions [10, 56, 57] and defects in solid-state systems such as nitrogen-vacancy (NV) centers [7, 12, 58–62], silicon-vacancy (SiV) centers [22, 63–66], quantum dots [67], and single rare-earth ions in solids [68, 69]. In particular, these solid-state platforms exhibit exceptional compatibility with advanced micro-and nano-fabrication techniques, paving the way for scalable and compact quantum network nodes and showing great prospects for large-scale integration.

Currently, development of quantum networks based on QMs is focused on establishing the heralded entanglement between distant QMs, which is a fundamental demonstration for quantum repeaters [9, 40, 43, 50, 51, 56, 59–62, 67, 70]. Additionally, these efforts are accompanied by the experimental validation of elementary functions of quantum networks, such as quantum state transfer, entanglement swapping, and nonlocal quantum gates across network nodes [10, 29, 30, 62, 71–75]. Current efforts are expanding toward long-distance entanglement and quantum operations across telecom optical fiber networks [7–12], which are critical for the integration of quantum networks with existing classical infrastructure.

In this Perspective, we present the fundamental requirements for establishing a quantum network based on QMs and analyze the experimental challenges and solutions encountered in long-distance demonstrations. We review state-of-the-art developments in remote quantum networks utilizing QMs based on various physical systems and discuss future efforts toward the construction of large-scale quantum networks.

## 2 The principle of quantum networks based on QMs.

Quantum networks can connect a large number of quantum nodes distributed in different regions, allowing quantum information to be transmitted and processed through quantum channels, as illustrated in Figure 1(a). Quantum entanglement, as a fundamental resource for many applications of quantum networks, must be distributed across quantum nodes efficiently. When two nodes are relatively close to each other, entanglement can be established by directly transmitting photons through optical fibers. In these basic applications, QMs can act as buffer devices to perform feedforward control via local operations and classical communication (LOCC) for some essential network functions, such as quantum state teleportation [75] and remote state preparation [76]. In simple two-node demonstrations with short distances, QMs can be replaced by fiber delay lines with fixed delay times [77, 78]. However, as communication distances expand to metropolitan scales and beyond, multiple nodes should be deployed between long-haul fiber links and the synchro-nization function of on-demand QMs would be indispensable. In addition, QMs can further offer versatile functionalities such as real-time manipulation of photonic qubits, waveform control, and pulse sequencer [79, 80].

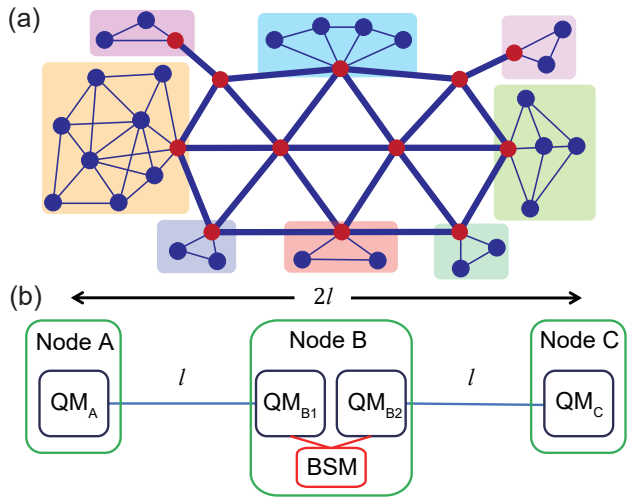

**Fig. 1:** Quantum networks based on QMs. (a) An artistic illustration of a quantum network. The red circles represent the long-distance repeater nodes, and the blue circles represent the user nodes, including quantum computing nodes, quantum sensing nodes, and so on. Different background colors indicate local networks. (b) A three-node quantum repeater. Each node contains QMs to store entanglement in the elementary links, which is then used for entanglement swapping through Bell-state measurements (BSMs) to extend the communication distance.

## 2.1 The role of QMs in quantum repeaters

Consider the simplest scenario involving three nodes, A, B, and C, as shown in Figure 1(b), where node C is the central node, and nodes A and B are located at the ends. The long-distance channel with a total length of $2l$ is divided into two shorter links, each with a length of $l$. The channel loss for directly transmitting quantum states from node A to node C is $2\eta_0 l$, where $\eta_0$ is the channel loss per unit distance (in dB). In the quantum repeater protocol [19], the first step is to attempt simultaneous entanglement distribution between nodes A and B, and between nodes B and C. However, due to the limited success probability of entanglement distribution, it is common for entanglement to succeed on one side but not the other. Here, we assume that the success probability of entanglement distribution, $p_0$, is the same for both the A-to-B and B-to-C links and depends solely on channel losses. The second step is the storage of the successfully created entangled states into QMs until both links, i.e., A-to-B and B-to-C, have successfully distributed entanglement. The third step is entanglement swapping [81] to generate entanglement between nonadjacent nodes A and C by implementing measurements in the node B. Without QMs, entanglement swapping can only be performed when entangled states are successfully distributed simultaneously between links A-to-B and B-to-C, which has a success probability of $p_0^2$, essentially the same scaling as direct transmission from node A to node C. With QMs, the success probability is determined by the inverse of the maximum time required for the two neighboring links to establish entanglement, which scales proportionally to $p_0$ [19, 21]. As a result, the total entanglement distribution time can increase only polynomially with channel distances, by repeating such process for many elementary links.

## 2.2 Light-matter entanglement between photons and QMs

To establish entanglement between distant quantum nodes, a prerequisite is the light-matter entanglement between flying photon qubits and stationary atomic QMs. For ensemble-based QMs, such entanglement could be generated either by using emissive QMs through Duan-Lukin-Cirac-Zoller (DLCZ) protocol [20] or using absorptive QMs to capture photons from external entangled light sources [50, 51]. For single-particle QMs, spin-photon entanglement can be generated by laser excitation and fluorescence detection in emissive QMs [9, 82]. For cavity-coupled single-particle systems, entanglement can be generated by absorbing external photons [54, 55] or through the spin-photon quantum logic operations [22, 29].

The entanglement generation rate of a quantum repeater strongly depends on the performances of QMs, including storage lifetime, efficiency, fidelity, bandwidth, and multimode capacity. There have been several elaborate reviews dedicated to the topic of QMs [52, 83–85], discussing these performances in detail. While individual metrics of QMs have achieved impressive performances in various systems, simultaneously meeting all these criteria in a single device remains a significant challenge for all candidate systems. Similar to classical networks, multiplexing is a key tool for enhancing the data rate in quantum networks [86, 86, 87]. For ensemble-based QMs, multimode stor-ages have been achieved with various degrees of freedom (DOF), such as temporal [88], spatial [89, 90], frequency [79, 91]. For single-particle-based QMs, it is also possible to generate multiplexed atom-photon interfaces by simultaneously addressing multiple atoms arranged in a single device [68, 92–94].

To facilitate long-distance transmission over standard telecom fiber, the flying photon should have wavelengths within the telecom band. However, except for a few systems, such as $Er^{3+}$ doped in solids [97, 98] and T-centers in silicon [96, 99], most QMs operate outside the telecom band. Currently, two primary approaches are being explored to resolve the incompatibility between operating wavelengths of QMs and telecom bands. The first approach is quantum frequency conversion (QFC), which converts the wavelength of photons to the telecom band through nonlinear processes while preserving their quantum properties. This process can be achieved by four-wave mixing in cold atomic ensembles [100] and frequency conversion in nonlinear crystals. The latter, due to its flexible tunability and higher conversion efficiency, has already been used to interface telecom-band photons with QMs based on cold atomic ensembles [8] and various single-particle systems [101, 102]. The second approach involves nondegenerate entangled photon pairs, where one photon is matched with QMs for long-lived storage, while the other one is in the telecom band for long-distance transmission. This method was initially implemented in REICs using light sources generated by spontaneous parametric down-conversion (SPDC) process [103, 104]. Spontaneous Raman scattering in cold atomic ensembles, serving as emissive QMs, could also enable the creation of polarization entanglement between 780-nm photons stored in the QM and telecom-band photons [105].

## 2.3 Heralded entanglement between QMs

The heralded entanglement between distant QMs can be treated as an elementary link for quantum repeaters [21]. The most common method for generating heralded entanglement is through entanglement swapping by implementing Bell-state measurement (BSM) in the intermediate node. A successful BSM heralds the generation of entanglement in an elementary link. This can be achieved either by the single-photon detection scheme [106, 107] or by the two-photon detection scheme [108–110].

The single-photon detection scheme operates similarly to a Mach-Zehnder interferometer [109] and is highly sensitive to phase fluctuations in the communication channel, which complicates practical implementation. However, it has the advantage of a faster entanglement establishment rate [60], as only one photon needs to reach the intermediate node for entanglement to be established. The success probability is proportional to $\sqrt{\eta}$ in the single-photon detection scheme instead of $\eta$ in the two-photon detection scheme, where $\eta$ denotes the transmittance of direct transmission through the elementary link. However, when dealing with number-state entanglement between ensemble-based QMs generated by single-photon interference, such as in the original DLCZ protocol [20], two parallel chains of number-state entanglement between QMs would be required to generate useful two-photon entanglement for practical applications [111], further increasing complexity and decreasing rate. It is worth mentioning that, with appropriate schemes, such as in some single-particle systems, single-photon detection scheme also provides an efficient solution to generate heralded two-particle entanglement directly [60]. On the other hand, the standard two-photon detection scheme, based on Hong-Ou-Mandel (HOM) interference, is more robust against phase fluctuations in communication channels. HOM interference requires only two indistinguishable photons with overlapping wave packets to interfere and does not rely on phase stability on a wavelength scale, making it more suitable for long-distance quantum communication [21, 109], especially in complex and harsh outdoor environments where maintaining long-term phase stability is challenging.

In addition to photon-photon interactions via interference, two alternative approaches can be utilized to generate heralded remote entanglement through cavity-enhanced atom-photon interactions, which exploit the effects of cavity quantum electrodynamics (CQED) [5, 112]. In the first method, strong coupling between a cavity and an atom enables the transfer of spin-photon entanglement from one node to another, where the emitted photons from one node can be efficiently and reversibly absorbed by the system in the other node, creating atom-atom entanglement [55, 113]. Successful absorption and entanglement creation can be heralded by nondestructive single-shot readout of the internal state of the atom or the detection of a scattered Raman photon in a crossed quantum channel [114]. The use of successive spin-photon quantum gates between an ancillary photon and two remote

**Tab. 1:** Overview of demonstrations of remote heralded entanglement between QMs.

| Physics system [a) | $\mathcal{D}_n/\mathcal{D}_c$ (km) [b) | Channel [c) | $\lambda_h$ [d) | $\mathcal{L}_i$ [e) | $\mathcal{F}$ or $\mathcal{C}$ [j) |
|---|---|---|---|---|---|
| | $\mathcal{L}_t/\mathcal{L}_f$ (dB) [f) | Scheme [g) | $\mathcal{R}_h$ or $[\mathcal{R}_B]$ [h) | $\mathcal{R}'_h$ or $[\mathcal{R}'_B]$ [i) | |
| Emissive QM-based | | | | | |
| NV centers [7] | 1.3/1.7 | Deployed | 637 nm | ~8 dB/km | $\mathcal{F} = (92 \pm 3)\%$ |
| | 15/15 | B | [0.00031 Hz] | [~0.0071 Hz] | |
| Quantum dots [67] | 0.002/- | Coiled | 968 nm | ~2 dB/km | $\mathcal{F} = (61.6 \pm 2.3)\%$ |
| | Negligible | A | 7300 Hz | 7300 Hz | |
| Cold-atomic ensembles [8] | 0.0006/22 | Deployed | 1342 nm | ~0.3 dB/km | $\mathcal{F} = (73.2 \pm 3.8)\%$ |
| | 17.6/8 | B | ~0.0067 Hz | ~0.042 Hz | |
| Cold-atomic ensembles [8] | 0.0006/50 | Coiled | 1342 nm | ~0.3 dB/km | $\mathcal{C} = 0.088(2)$ |
| | ~24.6/15 | A | ~1.5 Hz | ~8.4 Hz | |
| Single atoms [9] | 0.4/33 | Deployed | 1517 nm | ~0.2 dB/km | $\mathcal{F} = (62.2 \pm 1.5)\%$ |
| | ~13.5/8.6 | B | [0.012 Hz] | [~0.085 Hz] | |
| Single ions [95] | 0.23/0.52 | Deployed | 854 nm | ~3.5 dB/km | $\mathcal{F} = (88.0^{+2.2}_{-4.7})\%$ |
| | 4.0/4.0 | B | [0.058 Hz] | [~0.15 Hz] | |
| Cold-atomic ensembles [11] [k) | 12.5/19.7 | Deployed | 1342 nm | ~0.3 dB/km | $\mathcal{C} = 0.048(13)$ |
| | 15.0/8.2 | A | 0.83 Hz | ~2.14 Hz | |
| NV centers [12] | 10/25 | Deployed | 1588 nm | ~0.2 dB/km | $\mathcal{F} = (53.4 \pm 1.5)\%$ |
| | 27/10.8 | A | [0.022 Hz] | [~0.076 Hz] | |
| Silicon T-centers [96] | -/0.04 | Coiled | 1326 nm | 0.3 dB/km | $\mathcal{F} = (60 \pm 8)\%$ |
| | Negligible | B | [0.012 Hz] | [0.012 Hz] | |
| Absorptive QM-based | | | | | |
| REICs [50] | 0.0035/0.01 | Coiled | 880 nm | ~3.5 dB/km | $\mathcal{F} = (80.4 \pm 2.2)\%$ |
| | Negligible | B | [0.0003 Hz] | [0.0003 Hz] | |
| REICs [51] | 0.01/0.05 | Coiled | 1436 nm | ~0.3 dB/km | $\mathcal{C} = 0.0115(6)$ |
| | Negligible | A | 1430 Hz | 1430 Hz | |
| CQED QM-based | | | | | |
| SiV centers [65] | 0.006/35 | Deployed | 1350 nm | ~0.3 dB/km | $\mathcal{F} = (69 \pm 7)\%$ |
| | 29.7/17 | C | [0.00023 Hz] | [~0.012 Hz] | |

[a) Representative works on the heralded entanglement between remote QMs, which are categorized according to emissive, absorptive, and cavity quantum electrodynamics (QED) QMs, and listed in chronological order of publication. [b) $\mathcal{D}_n$ is the physical distance between the quantum nodes, $\mathcal{D}_c$ is the channel length between the quantum nodes. [c) The quantum channel types are divided into deployed fiber links and coiled fiber links wrapped inside the laboratory. [d) $\lambda_c$ represents the wavelength of the heralded photons through the channel. [e) $\mathcal{L}_i$ represents the ideal loss through the fiber links for the heralded photons. [f) $\mathcal{L}_t$ represents the total channel losses, including the loss caused by QFC, filters, and actual fiber links, and $\mathcal{L}_f$ represents the loss of actual fiber links. [g) Scheme represents different entanglement distribution schemes, where A is the single-photon detection scheme, B is the two-photon detection scheme and C is the spin-photon quantum logic gate based on CQED. [h) $\mathcal{R}_h$ ($[\mathcal{R}_B]$) represents the entanglement heralding rate (Bell-states generating rate). [i) $\mathcal{R}'_h$ ($[\mathcal{R}'_B]$) represents the estimated entanglement heralding rate (Bell-states generating rate) excluding the losses of fiber links. [j) $\mathcal{F}$ represents the fidelity of the heralded two-party entanglement, and $\mathcal{C}$ represents the average concurrence of the heralded entanglement (for the number-state entanglement case). [k) In ref. [11], among a total of three nodes, any two nodes are entangled with each other. Here, the shown data is from the two farthest nodes after the storage of 107 $\mu$s.

atom-cavity systems presents another viable option, enabling heralded entanglement creation [65], as well as quantum state and gate teleportation [29, 74] between remote nodes with local spin manipulations, photon detection, and active feed-forward control. The heralding photon detection of ancillary photons that successively reflect from two remote qubit modules signals the successful establishment of the teleportation and the remote entanglement. The cavity-based approach could also provide a potential solution to overcome the efficiency limitations of photonic BSMs [5, 115, 116].

Quantum repeaters can be categorized into three types based on the underlying QM technologies: emissive QM-based, absorptive QM-based, and CQED QM-based. A summary of recent demonstrations of heralded entanglement between remote QMs based on various platforms is provided in Table 1. Those works are selected primarily based on the channel distances, reflecting the system's potential to meet stringent experimental conditions in large-scale quantum networks. It is important to note that these demonstrations often involve trade-offs among key parameters. Generally, as the transmission distance increases, the experimental difficulty grows because of technical challenges such as longer storage times and greater channel losses and instabilities. In practice, achieving higher fidelity often comes at the expense of a lower entanglement rate [12, 96]. To better compare different systems and schemes, we introduce the parameters $\mathcal{R}'_h$ and $\mathcal{R}'_B$, which normalize the entanglement generation rate by exclude losses due to the physical fiber link. A more detailed discussion about these works is provided in the following chapter.

# 3 The state-of-the-art of quantum networks based on QMs.

In this section, we will introduce the state-of-the-art of building quantum networks based on QMs, with a focus on those systems that have already enabled implementations over metropolitan distances.

## 3.1 Cold atomic ensembles

Cold atomic ensembles have been widely explored for over two decades in quantum storage and the construction of quantum networks. Commonly used storage protocols include electromagnetically induced transparency (EIT) [42, 80, 105, 117–120], far off-resonant two-photon transition protocol (Raman scheme) [121, 122], and the DLCZ protocol [20]. The DLCZ protocol uses the spontaneous Raman scattering process after laser excitation to establish the number-state entanglement between photons and collective spin-wave excitation of atomic ensemble [20]. Combining two spontaneous Raman scattered fields from two spatial modes allows the entanglement between the spin-wave excitation and photonic polarization qubits to be efficiently generated [123], enabling two-photon detection scheme with other nodes. The DLCZ protocol, along with its variants, offers a promising pathway toward quantum repeaters based on emissive atomic ensemble [124–129].

C. W. Chou et al. generated the number-state entanglement between two atomic ensembles separated by 2.8 m based on the DLCZ protocol in 2005 [40]. Single-photon detection scheme is employed to verify the entanglement between two cesium ensembles. By generating pairs of such entanglement, i.e., four memories in total, to postselect a useful two-photon entanglement, the same group demonstrated functional quantum repeater nodes in 2007 [111]. In 2008, Z.-S. Yuan et al. extended this distance to 300 m of fiber (with an actual physical separation of 6 m) based on a two-photon detection scheme [130]. Various techniques have been employed to greatly enhance the readout efficiency and storage lifetime for large-scale applications, including putting the cold-atomic ensemble in a ring cavity for efficient retrieval [131], constraining the atomic motion using a three-dimensional optical lattice and using a magic compensating magnetic field. The recent record is an initial readout efficiency of $76 \pm 5\%$ with a $1/e$ storage lifetime of subseconds [41, 132]. In addition, efficient storage based on the EIT storage protocol is realized in an elongated cold atom system [133].

A major leap occurred in 2020 when Y. Yu et al. reported the heralded entanglement between two $^{87}$Rb atomic ensembles over 22 km of field-deployed fibers using two-photon detection scheme, as shown in Figure 2(a), and over 50 km of in-lab spooled fibers using single-photon interference, although the two QMs were still in the same laboratory [8]. QFC is employed to convert the emitted 795 nm photons to 1342 nm for long-distance fiber propagation. Although single-photon detection scheme is more technically challenging compared to two-photon detection scheme, it offers an enhancement of the entanglement generation rate of two orders in this demonstration. In 2024, J. Liu et al. successfully demonstrated a metropolitan quantum network consisting of three quantum nodes maximally separated by 12.5 km (Figure 2(b)) [11]. The heralded number-state entanglement can be generated between any two nodes using single-photon detection scheme based on active stabilization of the phase variance resulting from the deployed fiber and the control laser.

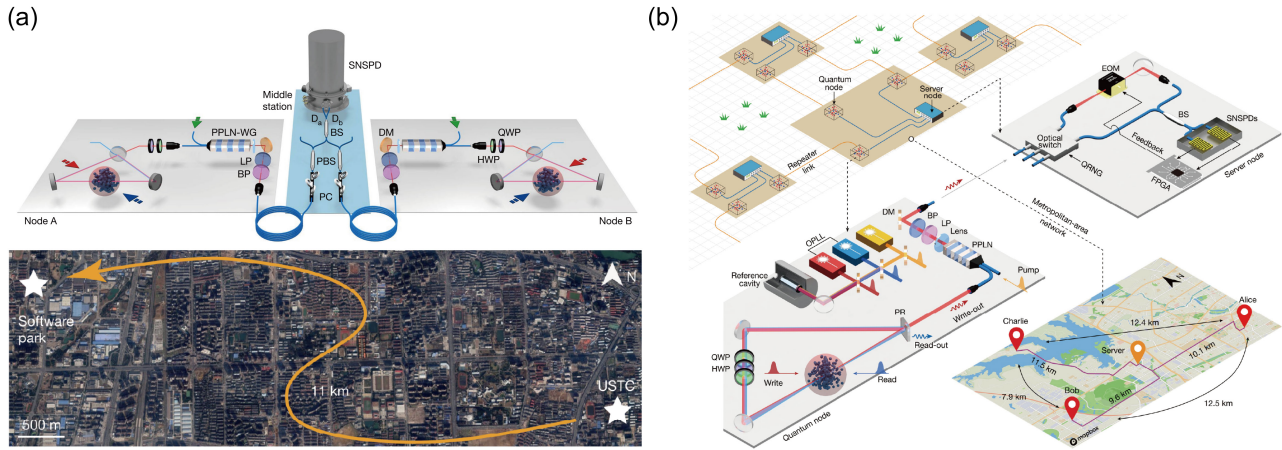

**Fig. 2:** The state-of-the-art demonstration of quantum networks based on cold-atom ensembles. (a) Remote heralded entanglement between two cold atomic ensembles over a 22 km urban fiber loop, while both atomic ensembles are in the same location. Reproduced with permission [8]. Copyright 2020, Springer Nature. (b) Heralded entanglement between any two QMs in three spatially separated nodes. Reproduced with permission [11]. Copyright 2024, Springer Nature.

It is worth noting that the atom-photon entanglement is generated based on a probabilistic single-photon emission process in the DLCZ protocol. As a result, there is a trade-off between the entanglement fidelity and the generation probability due to the multiphoton errors [20]. A potential solution is to employ the Rydberg interaction, which suppresses the probability of multiphoton excitation, allowing for the creation of deterministic atomic ensembles and photon entanglement with both emissive [134, 135] or absorptive schemes [136]. Integrating this scheme with long-lived spin-wave storage could be a promising approach to quantum networks based on cold atomic ensembles.

## 3.2 Single atoms and ions

Unlike atomic ensembles, qubits encoded in single particles can be precisely controlled and deterministically read out, enabling straightforward quantum computing processes. However, the efficient interface between single particles and single photons presents different experimental challenges.

Magneto-optical traps are typically used for pre-cooling to trap single neutral atoms, followed by optical dipole traps to confine individual atoms. In 2006, J. Volz et al. demonstrated the spin-photon entanglement between a single photon and a single trapped

$^{87}$Rb atom [137]. In 2007, T. Wilk et al. reported the realization of an atom-photon quantum interface based on an optical cavity, which is used to entangle a single $^{87}$Rb atom with a single photon, and subsequently map the quantum state of the atom onto a second photon [53].

In 2012, two research groups realized atom-atom entanglement between two separated single atoms with different approaches. Using cavity-based approaches, S. Ritter et al. demonstrated the transfer of an atomic quantum state and the creation of entanglement between two nodes with a distance of 21 m and linked by an optical fiber link of 60 m [55]. The entanglement is, however, not heralded, which can in principle be implemented by methods described in Section 2.3. Based on a two-photon detection scheme, J. Hofmann et al. demonstrated the heralded entanglement between two single independently trapped atoms separated by 20 m [139]. In 2022, as shown in Figure 3(a), by utilizing QFC with a system efficiency up to 57%, they increased the channel distance between two single atoms to 33 km linked by in-lab spooled telecom fibers, despite the physical distance of 400 m [9]. In 2024, after prolonging the coherence time of the single atom to 10 ms, they distributed the atom-photon entanglement by transmitting photons through spooled fibers with a length of 101 km, using the same QFC setup [140]. Building on cavity-assisted spin-photon quantum logic gates, quantum state teleportation [74] and quantum gate teleportation [29] were demonstrated over a distance of 60 m (see Figure 3(b)). The teleported CNOT gate was then used for the remote creation of all four Bell states between the two nodes [29]. It is important to highlight that the nonlocal quantum gate between remote nodes is also a fundamental component, indispensable for the realization of distributed quantum computing networks [31–36].

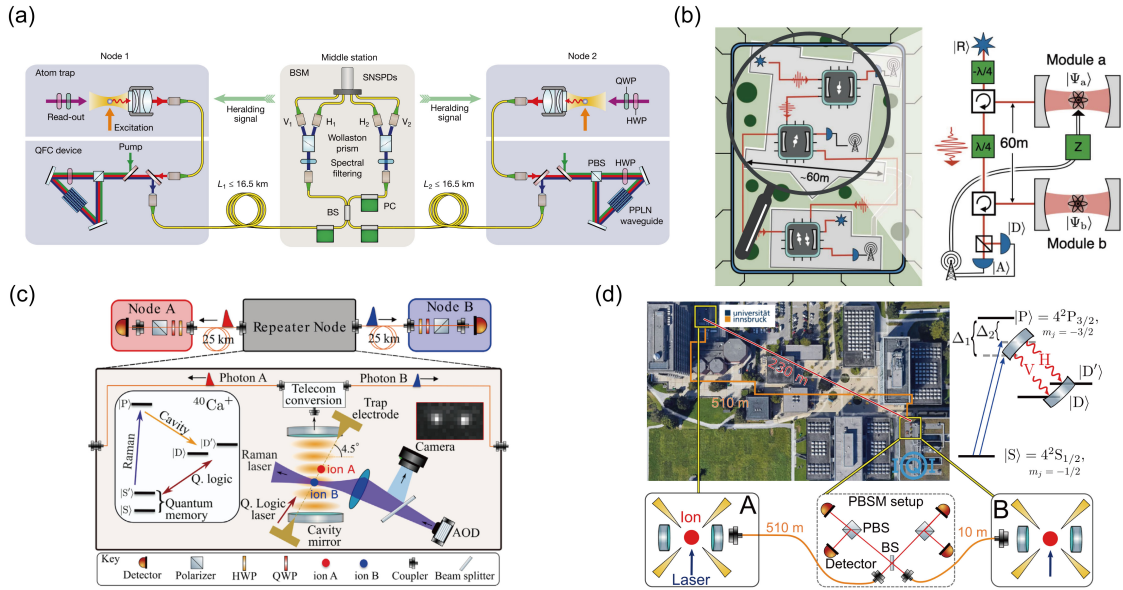

**Fig. 3:** The state-of-the-art demonstration of quantum networks based on single atoms and trapped ions. (a) Heralded entanglement between two single atoms over a 33-km coiled fiber. Reproduced under the terms of the CC BY 4.0 license [9]. Copyright 2022, Springer Nature. (b) A quantum-logic gate between two quantum nodes over a 60-m fiber link. Reproduced with permission [29]. Copyright 2021, AAAS. (c) A quantum repeater node based on trapped ions with an optical cavity enabled the establishment of photonic entanglement between two telecom-wavelength photons over 50 km of coiled fiber links. Reproduced with permission [138]. Copyright 2023, APS. (d) The heralded entanglement between two ions with a distance of 230 m over a 520-m fiber link. Reproduced with permission [95]. Copyright 2023, APS.

For single ions, radio-frequency Paul traps or Penning traps are employed, leveraging electric or magnetic fields to achieve precise confinement. The single ion qubit could exhibit a two-pulse spin echo coherence lifetime of 1.6 s and a spin coherence lifetime of 1 h upon protection with dynamical decoupling control [141], highlighting its significant potential for applications in long-distance quantum networks. The heralded entanglement between two trapped $^{171}$Yb$^+$ ions was first demonstrated in 2007 by two-photon detection [56], linked by fiber with a few meters. In 2020, by utilizing a novel collection geometry, the fidelity and generation rate of entanglement were significantly improved using two $^{88}$Sr$^+$ ions separated by 2 m [142]. The aforementioned work involves the emission of photons from ions in free space, single ions have also shown compatibility with high-finesse cavities in both weak and strong regimes [143, 144], making them another important candidate for cavity-enhanced quantum network nodes [5]. Nowadays, the distance was extended to 230 m with a 520-m optical-fiber link in 2023 (Figure 3(d)) [95]. In addition, the basic functions of quantum repeater node that distribute entanglement telecom-wavelength photons to two remote nodes over a 50 km coiled fiber link have been demonstrated (Figure 3(c)) [138]. The distance of the entanglement between light and trapped ions has increased from 50 km [145] to 101 km [93], and demonstrated in a 14.4 km urban fibers link [10]. These works lay the foundation for the entanglement between trapped ions at greater distances.

## 3.3 Defects in solids

Defects in solids, such as nitrogen-vacancy centers (NV) and silicon-vacancy centers (SiV) in diamond, exhibit optical and spin properties similar to those of single atoms but do not require sophisticated cooling and trapping techniques. The solid-state nature of these defects facilitates seamless integration with micro- and nanofabrication technologies, extending their potential to photonic and electronic platforms and promoting the realization of compact, stable, and highly functional quantum devices. In most cases, the integration of such defects with high-quality nanocavities or other nanostructures enhances their capabil-ity [146], enabling more efficient coupling to external optical channels and expanding their potential applications in quantum networks through cavity-enhanced effects [5, 147], while small-volume designs further enable more precise and efficient electrical and optical manipulation by creating stronger and more uniform control fields due to spatial confinement [69, 148, 149]. Moreover, this compatibility not only boosts device performance but also paves the way for large-scale integration, which is essential for the practical implementation of quantum networks and other advanced quantum technologies.

### 3.3.1 Nitrogen-vacancy centers

The NV center is a typical luminescent point defect in a diamond, where a nitrogen atom replaces a carbon atom, creating a vacancy simultaneously. This defect contains single electron spins with long spin coherence times [150]. Notably, nearby $^{13}$C nuclear spins close to $^{14}$N exhibit exceptionally long coherence times, reaching up to 75 s [151].

In 2010, the entanglement between a single photon and a single NV center was established by collecting photons using high numerical aperture objective lenses [152]. To improve the fluorescence collection of a single NV color center, researchers prepare a solid immersion lens on the surface using focused ion beam lithography [153], and apply this structure to demonstrate a two-qubit register [154]. In 2013, the entanglement distribution between two NV centers located 3 m apart was successfully demonstrated by electrical manipulation to align the zero-phonon lines of two NV centers [59]. The heralding was achieved through a double-heralded single-photon detection scheme, specifically the Barrett-Kok (BK) protocol [108], which can be classified as a two-photon detection scheme. In 2015, the NV center became the first physical system to establish entanglement between two QMs separated by a distance on the order of kilometers [7]. Heralded entanglement was established between two NV centers separated by 1.3 km using BK protocol (Figure 4(a)), enabling the loophole-free violation of a Bell inequality. However, the photons used for transmission were still in the visible spectrum, which imposes limitations on extending the entanglement distance. Later, with a single-photon detection scheme, P. C. Humphreys et al. further improved the entangling rate and suppressed the decoherence rate of electron spins through dynamical decoupling and then deterministically delivered a remote entanglement at every subsecond clock cycle [60].

Very recently, leveraging the technique of QFC and active phase stabilization across long fiber links, A. J. Stolk et al. have demonstrated the heralded entanglement between two NV centers spatially separated by 10 km and connected with 25-km deployed fiber [12]. The authors report a fidelity of $53.4 \pm 1.5\%$ for the heralded entanglement generation (Figure 4(b)). Although the use of single-photon detection scheme can enhance the entanglement distribution rate, there is an inherent trade-off between the fidelity of the state and the entanglement rate [60].

The primary factor limiting the scaling of NV centers over longer distances is the large phonon sideband, which results in a very low proportion of zero-phonon lines (ZPL) available for spin-entanglement generation, approximately 4% [155]. This limitation can be mitigated by embedding the NV center inside an optical microcavity, leveraging the Purcell effect to enhance both the emission rate and collection efficiency of photons at the ZPL. To prevent surface noise effects and fabrication-induced damage that contribute to frequency instability of the optical transition, an open Fabry-Pérot microcavity may provide a more effective solution compared to other nanophotonic structures fabricated at the diamond surface [156, 157].

### 3.3.2 Silicon-vacancy centers

Another defect in diamond, the SiV center, has also received widespread attention due to its excellent optical properties, which have high branching ratios of the ZPL, lifetime-limited fluorescence linewidth, and high brightness [158]. Both the nuclear spin memory based the $^{29}$Si [64] and the proximal $^{13}$C [63] exhibit long coherence times of approximately 1 s.

In 2016, A. Sipahigil et al. achieved strong coupling between artificially manufactured SiV color centers and photons by injecting Si$^+$ ions into diamond substrates with high-quality photonic crystal cavity [159]. Based on this integrated device, they demonstrated single-photon switch and entanglement of two SiV centers within a single nanophotonic device [159]. To enhance the photon collection efficiency, the same group developed a fiber-optical interface with coupling efficiency above 90% [160], enabling deterministic interface between photons and the electron spin memory based on SiV centers with coherence times exceeding 1 ms [63]. By utilizing efficient SiV-cavity cou-pling with large cooperativity, they further demonstrated the heralded spin–photon gate by nondestructive measurement of the spin states of the SiV center and demonstrated memory-enhanced quantum communication [22]. Based on the reflection-based spin–photon gate [5], nucleus–photon entanglement was successfully established using a photon–nucleus entangling gate, leveraging the strong coupling between the electron spin of SiV and the $^{29}$Si nuclear spin [64]. This work lays the foundation for subsequent demonstrations of the long-distance quantum repeater link.

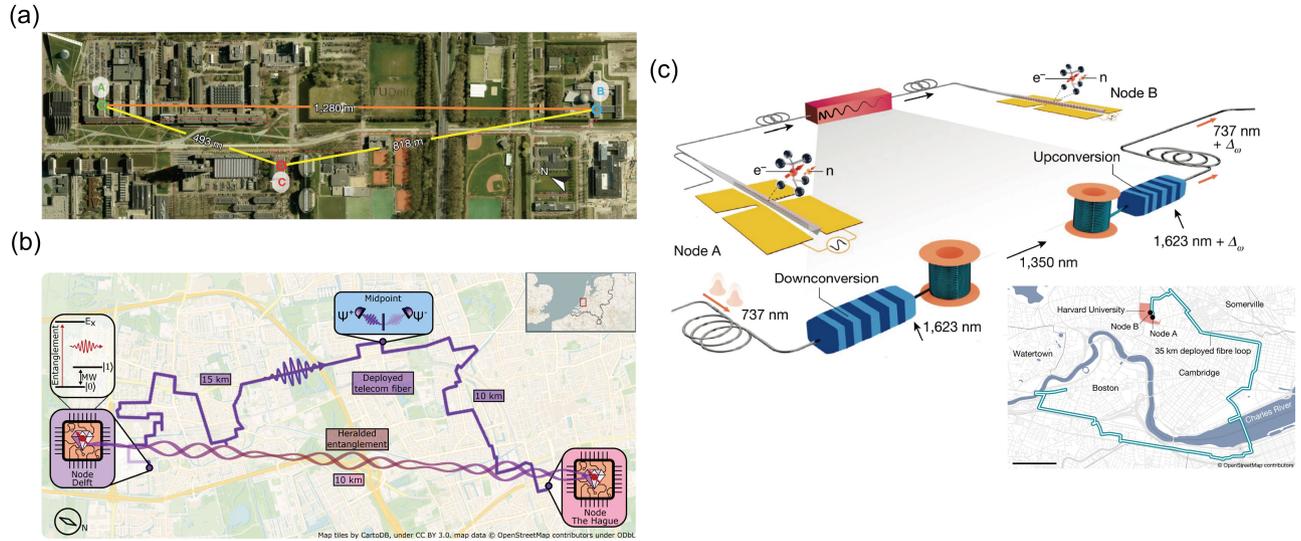

**Fig. 4:** The state-of-the-art demonstration of quantum networks based on defects in solids. (a) Remote heralded entanglement between two NV centers with a distance of 1.3 km. Reproduced with permission [7]. Copyright 2015, Springer Nature. (b) Heralded entanglement between two NV centers with a distance of 10 km over a 25-km urban fiber. Reproduced under the terms of the CC BY 4.0 license [12]. Copyright 2024, AAAS. (c) Heralded entanglement between two nearby SiV centers connected with a 35-km urban fiber link. Reproduced under the terms of the CC BY 4.0 license [65]. Copyright 2024, Springer Nature.

Building on this development in integrated nanophotonic nodes, in 2024, C. M. Knaut et al. demonstrated the establishment of heralded entanglement between the two $^{29}$SiV nodes through a 40-km low-loss fiber loop and a 35-km urban fiber loop (Figure 4(c)) [65]. The work established entanglement between a nuclear spin and a time-bin encoded photon at 737 nm at one node via a photon-nucleus entangling gate [64]. Subsequently, the 737-nm photons were upconverted to 1,350 nm, transmitted through long-distance optical fibers, and then downconverted back to 737 nm to interface with a second SiV center using the same gate. The detection of the 737-nm time-bin encoded photon heralded the entanglement between the two nodes. Although the two SiV nodes are located at the same laboratory, with a spatial separation of 6 m, this work highlights the potential of SiV-based quantum nodes for integration into large-scale quantum networks.

### 3.3.3 Other defect systems

In addition to color centers in diamond [161, 162] or other semiconductors [96, 163], defects in solids also encompass a wide range of different structures, including self-assembled semiconductor quantum dots, defects in various two-dimensional (2D) materials, nanotubes, and nanoparticles. These solid-state defects, typically at the nanoscale, exhibit distinctive optical and spin properties, making them well-suited for various quantum information processing applications. However, each system entails trade-offs between spin and photonic properties, and current research continues to explore competing systems, with no single system yet demonstrating a definitive overarching advantage. A more comprehensive discussion of the optical, spin, charge properties, material considerations, and potential applications of these defects, including their role in QMs, quantum light sources, and nanoscale sensing, can be found in references [164–167]. Here, we will discuss a few of these systems that have demonstrated the basic modules of quantum networks as described in Chapter 2, despite no long-distance demonstrations have been reported yet.

Among them, semiconductor quantum dots have garnered significant attention due to their unmatched high-brightness photon emission and excellent optical coherence properties [168–170]. Notably, quantum dots, when integrated with nanophotonic cavities or single-mode waveguides, have demonstrated high end-to-end single-photon source efficiencies, reaching up to 57% [171]. Moreover, these systems exhibit high indistinguishability and excellent entanglement fidelity in the emitted single photons or photon pairs [172–174]. As spin-based QMs, quantum dots have already demonstrated spin-photon entanglement and heralded entanglement between distant electron spins via photonic entanglement swapping [67, 175]. However, the limited electron spin coherence times of quantum dots, along with the absence of access to long-lived proximal nuclear spins, hinder their ability to meet the demands for memory qubits in long-distance quantum networks [168]. One possible approach is to couple to a nuclear spin ensemble in the host materials and exploit single nuclear spin-wave excitation as the memory qubit for storage [176]. Alternatively, a hybrid approach towards quantum repeater could involve interfacing quantum dots, which serve as deterministic single-photon or entangled photon sources, with absorptive QMs that support long-lived quantum storage [88, 177].

Recently, T centers in silicon hosts have emerged as a rapidly developing solid-state platform [178], offering an optical interface at the telecom O-band wavelength and compatibility with the mature nano-fabrication technologies of integrated silicon photonics. In 2024, F. Afzal et al. demonstrated high-quality spin-photon interfaces of T centers in isotopically enriched $^{28}$Si, and further realized the BK protocol-based generation of heralded spin–spin entanglement between two cavity-enhanced nodes located in separate cryostat [96], linked by 20 m of optical fiber, using the BK protocol. Utilizing the distributed entanglement, they also reported a non-local CNOT gate based on quantum gate teleportation, albeit in a postselected fashion. This progress positions T centers in silicon as one of the leading candidates for quantum networks without the need for QFC.

## 3.4 Rare-earth-ion-doped crystals

Rare-earth ions in solids can be treated as "a frozen gas of atoms", which is an important solid-state physical system for QMs. There are 17 types of rare earth elements, and their common feature is that the 4f electron layer is protected by fully arranged electron layers so that the inner 4f-4f transitions are insensitive to the environment and suitable for quantum storage [49, 179]. While the host crystal typically bring a relatively large inhomogeneous broadening for the rare earth ion ensemble, the homogeneous broadening, which is essentially the single-ion property, maintains a sharp spectral feature. As a prominent example, the hyperfine transition in the $^7F_0$ ground state of $Eu^{3+}$ in $Y_2SiO_5$ crystals exhibit a two-pulse spin echo coherence lifetime of 47 s and a spin coherence lifetime of 6 h under protection with dynamical decoupling control [180], which has enabled the coherent memory for light for 1 h [181]. In addition, due to the abundant variety of rare earth elements, the working wavelengths of rare earth ions are numerous, among which the $^4I_{15/2} \rightarrow {}^4I_{13/2}$ optical transition of $Er^{3+}$ ions is in the telecom $C$ band and has received widespread attention and research interests [98, 182–184].

Despite the long population lifetime of 4f-4f optical transitions, efficient optical detection of single isolated rare-earth ions has been achieved with the assistance of high-quality micro- and nanocavities to accelerate the optical relaxation through Purcell enhancement [68, 69, 185, 186]. Recently, the spin-photon entanglement has been generated between a telecom photon and the electron spin of a single $Er^{3+}$ ion in $CaWO_4$ interfaced with a Silicon-based photonic crystal cavity [187]. The heralded entanglement between two $Yb^{3+}$ ions, each placed in separate nanophotonic cavities fabricated directly from $YVO_4$ crystals, has also been successfully obtained through a single-photon detection scheme, although the two ions are within the same cryostat [188].

On the other hand, REICs are commonly used as ensemble-based absorptive QMs, which have the advantages of long coherence lifetimes [180, 181, 184], wide bandwidth [21, 183], large multimode capability [30, 79, 190] and ease of integration [52]. Typical quantum storage protocols implemented in REICs are atomic frequency comb (AFC) [39], gradient echo memory [191], and noiseless photon-echo (NLPE) [192]. In 2021, two research works used external memory-compatible sources, combining multiplexed AFC storage to demonstrate heralded entanglement between two absorptive QMs [50, 51]. The distance between the two QMs is 3.5 m in Ref. [50] (Figure 5(b)) and 10 m in Ref. [51] (Figure 5(a)) respectively. To enable long-distance applications [30, 75, 189, 190, 193], SPDC process is employed to generate entanglement between a memory-compatible pho-ton and a telecom-band photon. Based on a cavity-enhanced SPDC source and a $Pr^{3+}$:$Y_2SiO_5$ crystal, the multiplexed quantum teleportation with a 1-km coiled fiber is demonstrated in a local lab [75], and the distance of light-matter entanglement reaches 50 km based on urban fiber cables despite the memory lifetime is still below the photonic propagation times (Figure 5(c)) [189].

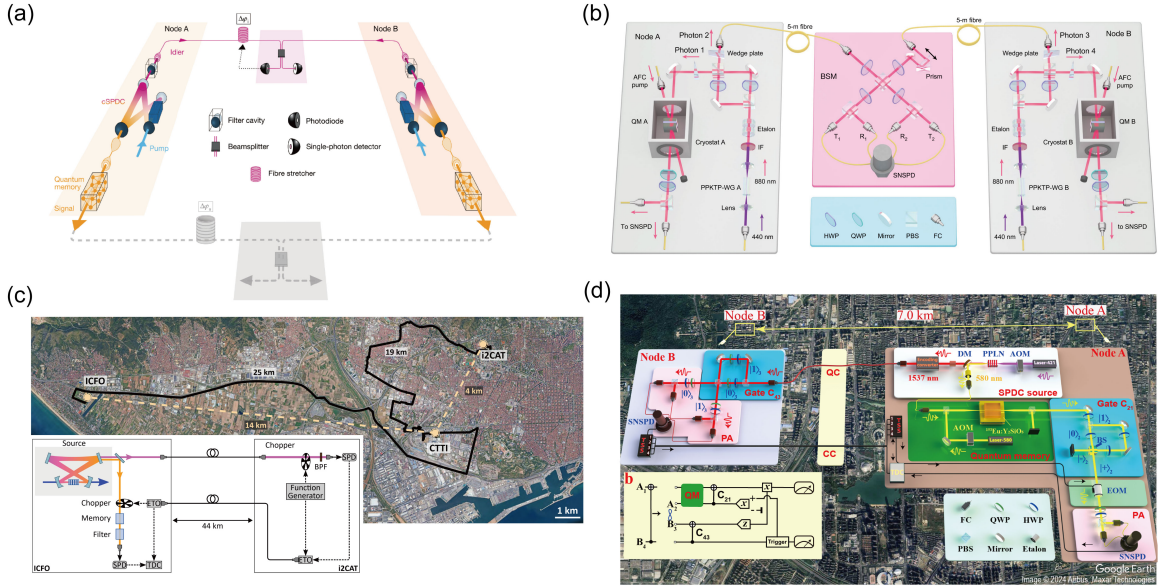

**Fig. 5:** The state-of-the-art demonstration in quantum networks based on rare-earth-ion-doped crystals. (a) Telecom-heralded single-photon entanglement between two multiplexed absorptive QMs over a 50-m fiber link based on $Pr^{3+}$:$Y_2SiO_5$ crystals. Reproduced with permission [51]. Copyright 2021, Springer Nature. (b) Heralded two-party entanglement between two multiplexed absorptive QMs over a 10-m fiber link based on $Nd^{3+}$:$YVO_4$ crystals. Reproduced with permission [50]. Copyright 2021, Springer Nature. (c) Transmission of light-matter entanglement with a distance of 44 km. Reproduced under the terms of the Optica Open Access Publishing Agreement [189]. Copyright 2023, Optica. (d) Distributed photonic quantum computing over 7.0 km based on the entanglement between a telecom photon and a $^{153}Eu^{3+}$:$Y_2SiO_5$ crystal. Reproduced under the terms of the CC BY 4.0 license [30]. Copyright 2024, Springer Nature.

Recently, our group demonstrated distributed photonic quantum computing between two nodes separated by 7.0 km and connected with a fiber of 7.9 km (Figure 5(d)) [30]. A pair of entangled 580-nm and 1,537-nm photons is distributed between two nodes and two two-qubit controlled gates are implemented in both nodes through path and polarization encoding on each photon. A $^{153}Eu^{3+}$:$Y_2SiO_5$ crystal is employed to multiplexed storage of the 580-nm photons to wait the classical communication from the distant node and to enable the local operation depending on the classical communication results, to complete the quantum gate teleportation protocol [194, 195]. In addition, the Deutsch-Jozsa algorithm and the quantum phase estimation algorithm are implemented in these two remote nodes based on these teleported nonlocal gates. This is the first proof-of-principle demonstration of distributed quantum computing networks over a metropolitan scale, laying out the foundations for scalable quantum computing based on large-scale quantum networks.

Similar to the DLCZ protocol, the SPDC process generates entanglement in a probabilistic manner. Since the absorptive QMs are flexible with the choice of quantum light sources, this problem could be overcome with a deterministic photon source to replace the SPDC source [88]. Going to much longer distances can be achieved by using spin-wave-based quantum storage protocols such as the full AFC protocol [39] and the NLPE [192]. Spin-wave quantum storage has enabled the storage of time-bin qubits for 20 ms in RE-ICs [196], with a potential lifetime of hours [180, 181], which could further enable the construction of transportable QMs [197, 198] for flexible long-distance quantum communication without the need of deployed fiber channels. Implementation of long-lived absorptive QMs and single-ion qubit based on $Er^{3+}$ ions could directly support the telecom photonic interface [68, 184, 187], making it another promising candidate system for applications in future quantum networks.

# 4 Conclusion and Prospective

With the rapid development of each physical system, memory-based quantum network demonstrations are moving toward greater distances and enhanced practicality. Current progress has successfully established long-distance entanglement between two QMs. Notably, heralded entanglements between quantum nodes over 10 km have been demonstrated in cold-atom ensembles [11], and NV centers [12] very recently, representing an important milestone for large-scale quantum networks, although the nonlocality is not revealed in these two works due to limited rates and fidelity.

To further increase the distance and entanglement distribution rate, advancements are required in several key aspects. First, optimizing the overall performances of QMs, including fidelity, device efficiency, lifetime, and multimode capacity, is crucial. The demonstration of long-distance quantum repeaters with high-quality entanglement, i.e., heralded atomic entanglement with nonlocality over long distances, will mark the next significant milestone for quantum networks. Second, it is equally crucial to develop high-brightness, memory-compatible quantum light sources [173, 199] for absorptive QMs or generate atom-photon entanglement with high efficiency and high fidelity. For the first two aspects, optical cavities and waveguides provide general tools to enhance light-matter interactions and optimize the efficiency of both quantum light sources [200, 201] and quantum storage systems [42, 52, 132, 132, 202]. In the meantime, ongoing efforts have been made to preserve the intrinsic spin and optical properties of quantum systems during integration and fabrication process [52, 203].

Lastly, advancements in other enabling technologies are also important. These include the development of single-photon detectors with higher efficiency, lower dark counts, and photon-number-resolving capability, as well as the improvement in quantum channels. For most physical systems, QFC is crucial for converting photon wavelength to achieve minimal losses in quantum channels, but the inefficiency and noise of QFC impact the rate and fidelity of entanglement distribution, especially for multinode applications. Recent breakthroughs in hollow fibers present a potential solution for low-loss photon transmission across a wide range of wavelengths [204, 205]. However, current challenges such as high attenuation and large costs persist.

Realizing practical quantum networks based on QMs is a multifaceted challenge that requires the ad-vancement of comprehensive quantum technologies. With continuous developments, it should be possible to demonstrate the elementary link of quantum repeaters with distances of hundreds of km in the coming years. However, for large-scale quantum networks, the next step after heralded entanglement distribution would be the connection of different elementary links, which involves at least four QMs arranged in order in three cascaded nodes, as illustrated in Figure 1(b). This progression demands a substantial enhancement in the entanglement distribution rate within each elementary link, as well as efficient cascading between neighboring QMs [43, 138]. Only by achieving these capabilities can the advantages of quantum repeaters over direct fiber transmission be fully validated. Nevertheless, with the rapid progress in quantum information technologies and sustained research efforts across diverse matter systems, the realization of this goal is within reach in the foreseeable future.

**Acknowledgment:** Z.-Q. Z acknowledges the support from the Youth Innovation Promotion Association CAS.

**Research funding:** This work is supported by the Innovation Program for Quantum Science and Technology (No. 2021ZD0301200), the National Natural Science Foundation of China (Nos. 12222411, 11821404, 12204459 and 12474367), and the China Postdoctoral Science Foundation (2023M743400).

**Author contributions:** All authors have accepted responsibility for the entire content of this manuscript and approved its submission.

**Conflict of interest:** Authors state no conflict of interest.

**Data availability statement:** Data sharing is not applicable to this article as no datasets were generated or analyzed during the current study.